\documentclass[conference]{IEEEtran}
\IEEEoverridecommandlockouts

\usepackage{cite}
\usepackage{amsmath,amssymb,amsfonts}
\usepackage{algorithmic}
\usepackage{graphicx}
\usepackage{textcomp}
\usepackage{xcolor}
\usepackage{booktabs}
\usepackage{multirow}
\usepackage{hyperref}
\usepackage{subcaption}
\def\BibTeX{{\rm B\kern-.05em{\sc i\kern-.025em b}\kern-.08em
    T\kern-.1667em\lower.7ex\hbox{E}\kern-.125emX}}
\begin{document}

\title{Zero-Shot Lombard Speech Synthesis with Controllable Style Embeddings

}

\author{\IEEEauthorblockN{Seymanur Akti}
\IEEEauthorblockA{\textit{Karlsruhe Institute of Technology (KIT)} \\
\textit{KIT Campus Transfer (KCT)}\\
Karlsruhe, Germany \\
seymanur.akti@kit.edu}
\and
\IEEEauthorblockN{Alexander Waibel}
\IEEEauthorblockA{\textit{Karlsruhe Institute of Technology (KIT)} \\
Karlsruhe, Germany \\
\textit{Carnegie Mellon University (CMU)}\\
Pittsburgh, USA \\
alexander.waibel@kit.edu}

}

\maketitle

\begin{abstract}
The Lombard effect plays a key role in natural communication, particularly in noisy environments or when addressing hearing-impaired listeners. We present a controllable text-to-speech (TTS) system capable of synthesizing Lombard-like speech in a zero-shot manner without requiring Lombard-specific training data. Our approach extends F5-TTS with a learned style embedding representation and analyzes the resulting latent space using principal component analysis (PCA) to identify directions associated with Lombard-related attributes. By manipulating these directions, we obtain interpretable control over vocal effort and articulation and generate speech at different Lombard levels. Experimental results show that the proposed method preserves speaker identity and naturalness, improves intelligibility under noisy conditions, and generalizes to previously unseen speakers. These findings demonstrate that style-embedding manipulation provides an effective and scalable framework for controllable zero-shot Lombard speech synthesis.

\begin{IEEEkeywords}
    speech synthesis, lombard speech
\end{IEEEkeywords}

\end{abstract}

\section{Introduction}
\label{sec:intro}

Humans naturally adapt their speaking style under noisy conditions or when addressing listeners with hearing difficulties, a phenomenon commonly referred to as the Lombard effect. Compared to neutral speech, Lombard speech exhibits distinct acoustic characteristics, including increased vocal energy, elevated and expanded fundamental frequency (F0), reduced spectral tilt, slower speaking rate, and enhanced articulation, often resulting in greater diversity in the formant space~\cite{junqua1993lombard, soltau2000specialized}. These adaptations occur naturally in human speech under challenging listening conditions, with the goal of improving intelligibility. Earlier studies have shown that such acoustic differences can also affect automatic speech recognition (ASR) performance, motivating the development of adapted recognition models~\cite{soltau2005compensating, soltau2002compensating}.

From a speech synthesis perspective, simulating these adaptations in text-to-speech (TTS) systems is important for applications such as hearing-assistive technologies, robust speech synthesis in noisy environments, and error-repair strategies in interactive systems~\cite{suhm1999model, constantin2022interactive}. However, despite recent advances in expressive and controllable speech synthesis, relatively little work has focused on controllable TTS systems that adjust speech according to listener needs or environmental conditions.

In this work, we formulate Lombard speech synthesis as an interpretable latent-space control problem. We adopt F5-TTS~\cite{chen2024f5} and incorporate style embeddings learned from a large, prosodically diverse dataset. Rather than learning an explicit plain-to-Lombard mapping, we analyze the learned style space using principal component analysis (PCA), following prior work on interpretable style control~\cite{sorin2020principal, nishihara2024low}. We identify latent directions associated with Lombard-related speech characteristics and manipulate them by shifting the corresponding principal components (PCs) and applying the inverse PCA transformation, enabling zero-shot Lombard speech synthesis of unseen speakers without Lombard samples from the target speaker. Although labeled vocal-effort and articulation data are used to identify these directions, they are not used to train or fine-tune the TTS model. This post-training formulation could potentially be extended to other style characteristics represented in the learned latent space. In addition, we incorporate inference-time duration control to adjust speech length according to the desired speaking rate, which is particularly important for simulating clear speech.

We further validate that shifts along the identified principal components produce systematic and controllable changes in key acoustic correlates of Lombard speech, including spectral energy, F0 variance, spectral tilt, and speaking rate. Our framework provides both separate and joint controls for vocal effort and articulation, enabling zero-shot Lombard-like speech synthesis for unseen speakers without requiring speaker-specific Lombard recordings. Experiments evaluate the effects of Lombard control on speaker similarity, ASR performance under noisy conditions, and acoustic correlates of vocal effort and articulation. The results show stable speaker similarity across Lombard levels and improved recognition robustness as Lombard control strength increases. Audio samples are available on our demo page.\footnote{\href{}{https://seymanurakti.github.io/lombard-speech/}}

\begin{figure*}[!]
\centering

\includegraphics[width=\linewidth]{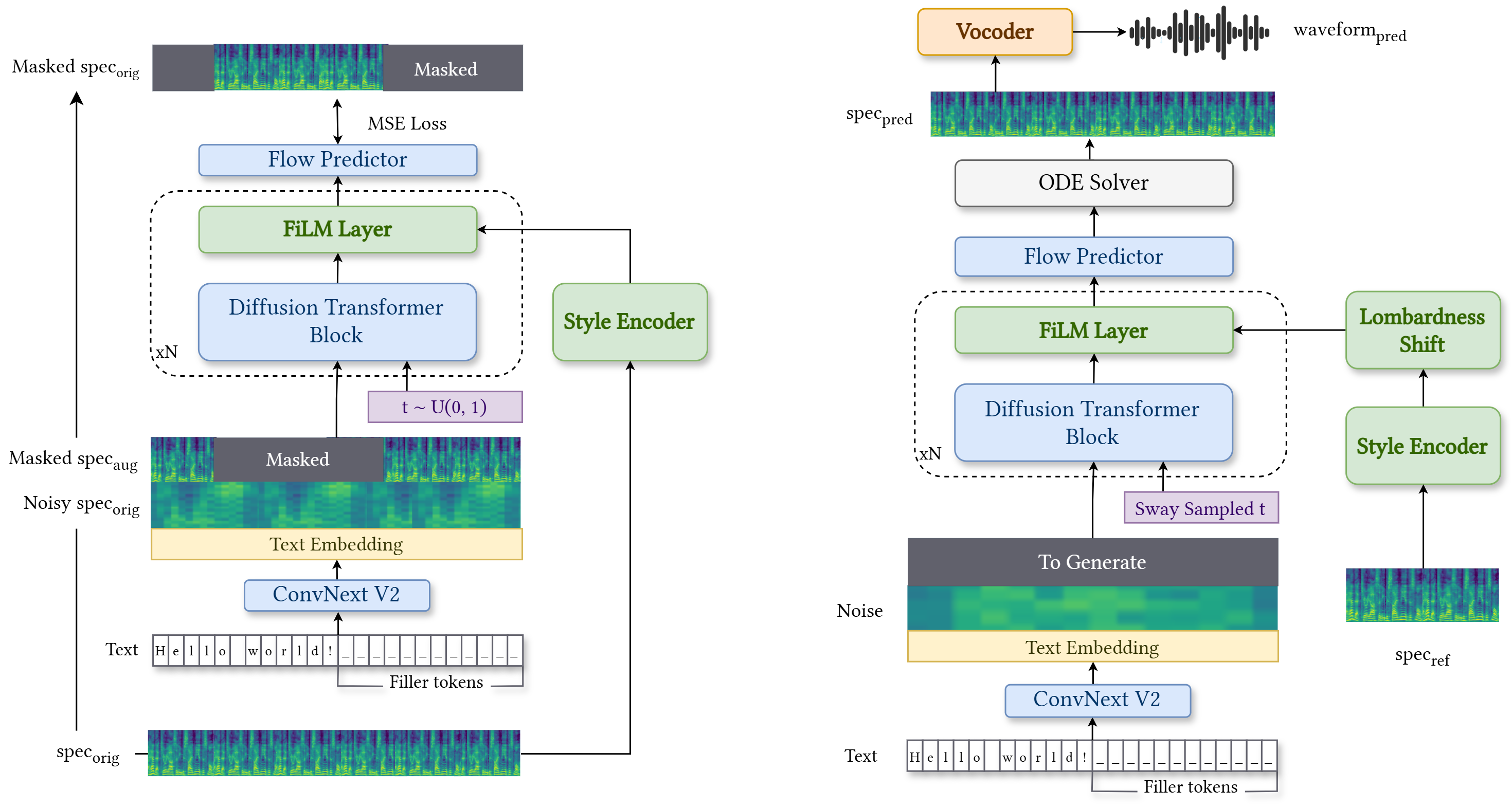}
\caption{Training (left) and inference (right) pipelines of the proposed system. Blue blocks represent components inherited from the original F5-TTS model, while green blocks indicate modules introduced in the proposed version.}

\label{fig:architecture}

\end{figure*}

\section{Related Work}

Many previous approaches treated Lombard speech production as a speech enhancement problem, where regular speech is transformed into Lombard-like speech by modifying spectral features and redistributing energy~\cite{brouckxon2008time, sauert2012near, taal2013sii, taal2014speech}. These include non-linear processing such as high-pass filtering~\cite{niederjohn2003enhancement}, dynamic range compression~\cite{zorila2014spectral,kandia2012speech}, manipulation of relative consonant energy ratios~\cite{skowronski2006applied}, and time-domain modifications to mitigate overlaps between speech and noise~\cite{aubanel2013information, verhelst2000overlap}.

In addition to signal processing-based methods, several works have explored Lombard speech synthesis within text-to-speech (TTS) frameworks. These include HMM-based modeling of Lombard or hyper-articulated speech~\cite{raitio2011analysis, suni2013lombard, picart2014analysis}, fine-tuning TTS models on Lombard speech~\cite{bollepalli2019lombard, hu2021whispered}, and formulating Lombard synthesis as a style conversion problem~\cite{bollepalli2019normal, paul2020enhancing}. Other approaches control spectral tilt to vary vocal effort in TTS~\cite{raitio2022vocal}, or dynamically adapt synthesis parameters based on SNR feedback~\cite{novitasari2021dynamically}. More recent methods leverage controllable style embeddings to modify articulation via principal component analysis (PCA)~\cite{nishihara2024low}, or train a joint control mechanism for vocal effort and hyper-articulation using small amount of data~\cite{akti2026synthesizing}.

Many Lombard-specific TTS approaches rely on Lombard speech data during training, either through explicit supervision or parallel Lombard/plain recordings. This reliance can make adaptation to unseen speakers or conditions dependent on the availability and diversity of Lombard training data. The method most similar to ours~\cite{lobato2025gradual} introduces zero-shot, speaker-adaptive Lombard control by learning a mapping between plain and Lombard speaker embeddings and interpolating between them during inference. However, speaking-rate control was implemented through linear time stretching, whereas our framework combines latent-space control of Lombard-related vocal effort and articulation with inference-time duration control. In contrast to approaches that learn an explicit plain-to-Lombard transformation, we identify attribute-related directions in a learned style space and manipulate these directions at inference time. This provides a speaker-agnostic mechanism for zero-shot Lombard-like synthesis while allowing both separate and joint control of vocal effort and articulation.

\section{Methodology}
\label{sec:methodology}

We adopt F5-TTS~\cite{chen2024f5} as the baseline for controllable Lombard speech synthesis due to its high synthesis quality and built-in duration control. F5-TTS is a flow-matching-based TTS model that, during training, receives masked mel-spectrograms together with noisy mel-spectrograms and text embeddings. Text embeddings are padded at the end to match the sequence length of the target mel spectrogram. The model then implicitly learns alignment between characters and mel-spectrogram frames during training, eliminating the need for an external duration predictor or forced alignments. At inference, the original model takes a reference audio sample and its transcription, using an in-context learning strategy to generate speech for new text conditioned on the reference signal. This conditioning mechanism allows the model to capture speaker identity, prosody, and speaking style directly from the reference utterance, contributing significantly to its synthesis quality.

This design has several limitations: it requires reference transcriptions at inference time and may suffer from cross-lingual mismatch when the reference audio is in a different language, potentially introducing accent drift or artifacts. Furthermore, because speaker identity, linguistic content, and speaking style are jointly represented through the reference utterance, direct manipulation of individual style attributes is difficult. To address these issues, we remove the need for reference text during inference and instead inject paralinguistic information via a fixed-size style embedding. This modification is particularly important for Lombard control, as it enables direct manipulation of style representations during synthesis. Figure~\ref{fig:architecture} shows overall architecture of the proposed system.

\begin{figure*}[!]
    \centering

    \begin{subfigure}[t]{0.48\textwidth}
        \centering
        \includegraphics[width=\linewidth]{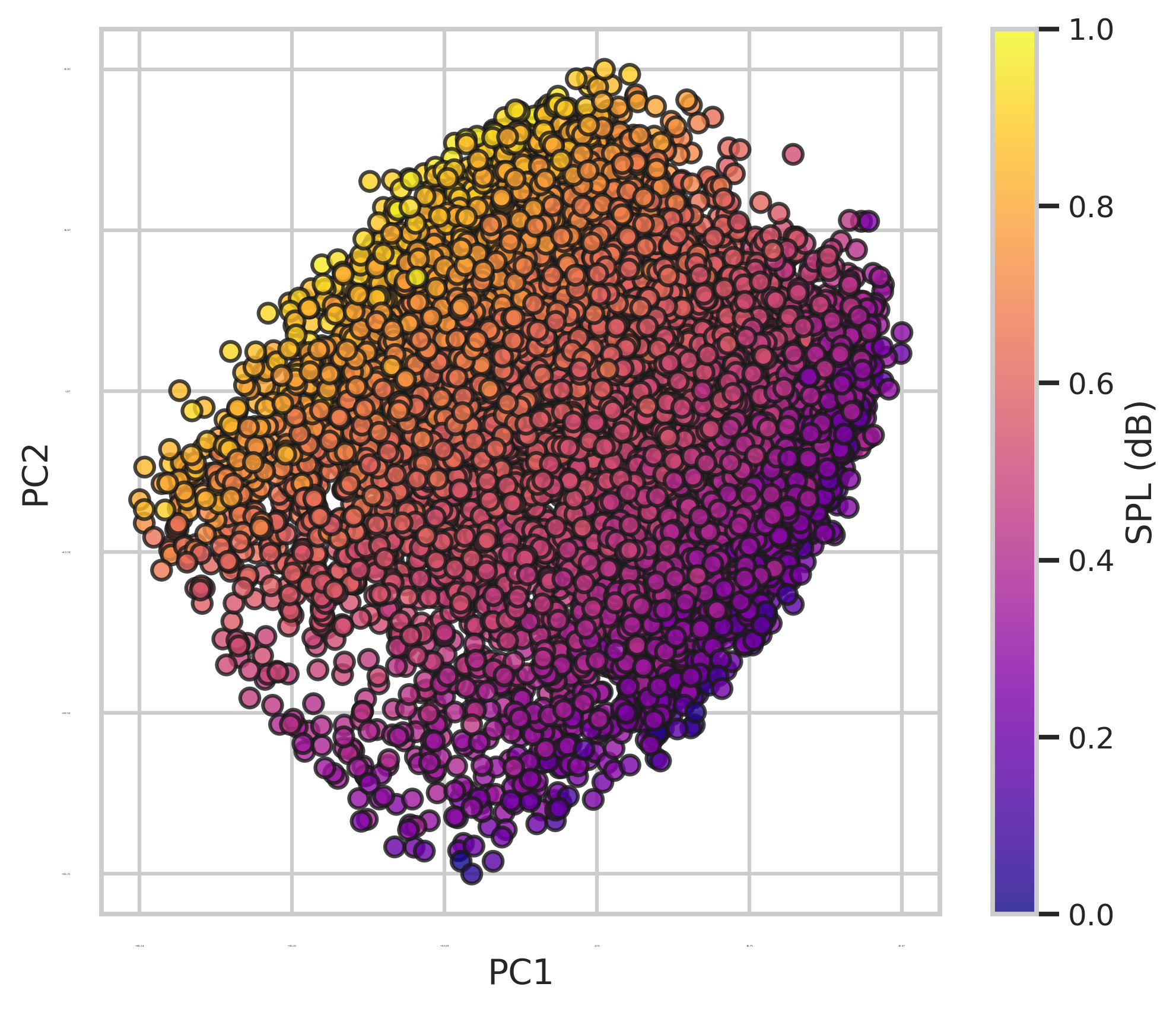}
        \caption{Correlation with vocal effort in the AVID embeddings space.}
        \label{fig:pcs_loudness}
    \end{subfigure}
    \hfill
    \begin{subfigure}[t]{0.48\textwidth}
        \centering
        \includegraphics[width=\linewidth]{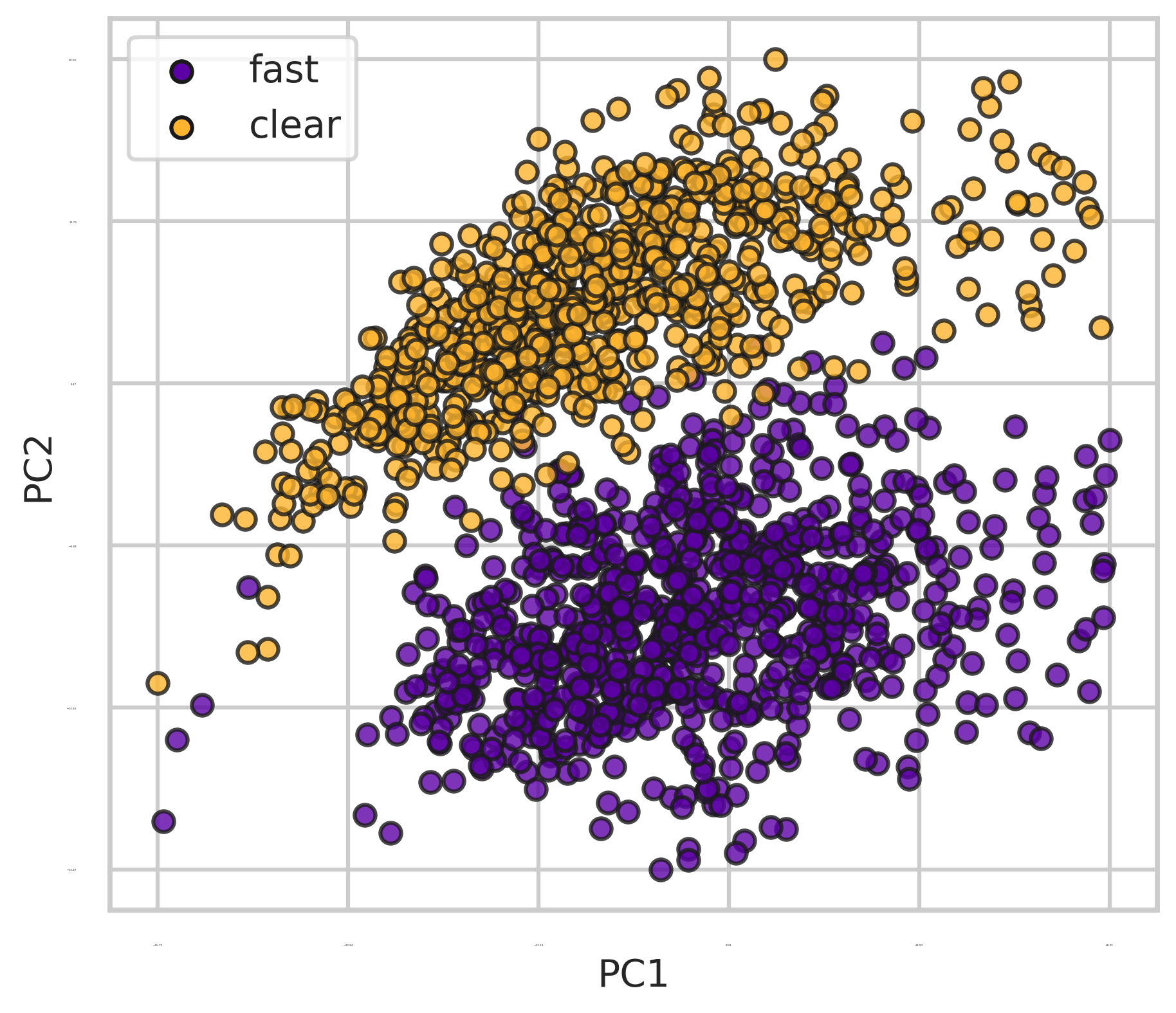}
        \caption{Correlation with articulation in ALBA embeddings space.}
        \label{fig:pcs_clarity}
    \end{subfigure}

    \caption{PCA analysis of style embeddings.}
    \label{fig:pca_analysis}
\end{figure*}

\subsection{Model Adaptation}


The F5-TTS base model consists of 22 Diffusion Transformer (DiT) blocks. Preliminary experiments indicated that the early layers are primarily responsible for learning phoneme-to-frame alignment, whereas speaker and speaking-style characteristics are represented more strongly in later layers. Accordingly, we freeze the first two DiT blocks to preserve the alignment capabilities learned by the base model while adapting the remaining layers for style control.

Our approach does not rely on the in-context style transfer mechanism of the baseline model. Rather, we inject style information through FiLM conditioning~\cite{perez2018film} at the unfrozen DiT blocks, where intermediate activations are scaled and shifted using parameters derived from fixed-size style embeddings. The style embeddings used for conditioning are extracted by a style encoder with ECAPA-TDNN architecture~\cite{desplanques2020ecapa} which is trained from scratch jointly with the TTS model. The encoder learns a 1024-dimensional embedding space that captures paralinguistic information, including speaker identity and expressivity, thereby replacing the baseline's in-context conditioning pathway. This design provides a continuous latent space, which we later exploit to control Lombard-related attributes.

To encourage the model to rely on the style embeddings for speaker identity and speaking style, rather than implicitly extracting this information from the masked mel-spectrogram inputs, we augment the masked mel-spectrograms with controlled formant shifts during training. These perturbations partially suppress speaker-specific cues in the acoustic input, forcing the model to recover them through the style embedding conditioning in order to reconstruct the spectrogram. As a result, the style embeddings become a stronger and more disentangled source of speaker and style information.

During inference, the model generates speech conditioned solely on the input text and a style embedding extracted from a short reference audio sample, eliminating the need for a reference transcription.

\subsection{Lombard Effect Control}

To enable controllable Lombard-like synthesis, we manipulate the style embeddings extracted from the style encoder. The style encoder is trained jointly with the TTS model on the English subset of the Emilia dataset~\cite{he2024emilia}, which contains a diverse range of speakers and speaking styles, with recordings collected in the wild. During training, the style encoder learns to represent a wide variety of speech characteristics, allowing the embedding space to capture substantial variation in both prosodic and speaker-related features. We hypothesize that this space is sufficiently expressive to also encode attributes related to vocal effort and speaking rate, even though it is not explicitly supervised for these properties.

This becomes more evident when the latent space of the style embeddings is analyzed using principal component analysis (PCA). We perform PCA separately on style embeddings from the speech data with (1) varying vocal effort and (2) varying articulation. The visualization at Figure~\ref{fig:pca_analysis} shows that it is possible to identify certain directions in the latent space, strongly correlated with the respective attributes.

\textbf{Vocal Effort PCA} is performed on style embeddings extracted from the AVID dataset~\cite{alku2024avid}, which contains speech produced at multiple vocal effort levels (soft, normal, loud, and very loud), along with calibrated speech pressure level (SPL) labels for 50 speakers in total. We calculate Pearson correlation between the principal components and the SPL labels to identify most strongly correlated components with vocal effort, which are then used for latent-space manipulation. PCA is computed using embeddings from the first 40 speakers (20 male and 20 female) in the dataset.

\textbf{Articulation PCA} is performed on style embeddings extracted from the ALBA dataset~\cite{valentini2019alba}, which contains fast, normal, and clearly articulated speech produced by a single female speaker. The principal components exhibiting the strongest separation between articulation classes are selected as control directions for articulation manipulation.

Let $z \in \mathbb{R}^{d}$ denote a style embedding extracted from reference speech. Let $P_{E}$ and $P_{A}$ denote the PCA projection matrices for vocal effort and articulation, respectively, and let $e_{E}$ and $e_{A}$ denote the corresponding principal axes for each attribute. Sequential manipulation is performed as follows:
\begin{align}
z' &= P_{E}^{-1} \big( P_{E}(z) + \alpha_{E} e_{E} \big), \\
z'' &= P_{A}^{-1} \big( P_{A}(z') + \alpha_{A} e_{A} \big),
\end{align}
where $\alpha_{E}$ and $\alpha_{A}$ are user-specified coefficients controlling vocal effort and articulation, respectively. By modifying the embedding along the attribute-specific directions, the model provides separate controls for vocal effort and articulation while retaining the speaker characteristics encoded by the reference embedding. This design enables synthesizing Lombard speech for speakers that were not used to estimate the attribute-specific PCA directions.

Finally, we adjust the total duration of the generated speech by modifying the syllable-to-time ratio, leveraging the built-in duration control of F5-TTS to further enhance intelligibility and perceived clarity. Unlike linear time-stretching, this mechanism does not uniformly scale the spectrogram. Instead, F5-TTS dynamically adjusts phoneme-to-frame alignments during inference to match the target duration, enabling non-linear modification of acoustic and linguistic timing.

\section{Experiment and Results}
\label{sec:experiments}

We first evaluated our model against the F5-TTS baseline on multilingual LibriSpeech test sets~\cite{Pratap2020MLSAL} for evaluating the style transfer capability under two scenarios: (1) using English prompt audio and (2) using German prompt audio. This allowed us to assess both monolingual English and cross-lingual speaker adaptation. The evaluation metrics are word error rate (WER) from Whisper-large-v3\footnote{https://github.com/openai/whisper}, speaker similarity (SSIM) computed as the cosine similarity between embeddings from a speaker verification model\footnote{https://huggingface.co/microsoft/wavlm-base-sv}, and UTMOS~\cite{saeki2022utmos} for predicting perceived naturalness.

For evaluating Lombard speech, we used the 10 AVID speakers (5 male and 5 female) that were held out from the PCA analysis. In total 500 samples were evaluated where each speaker has 50 samples.  Ground truth Lombard levels in the AVID dataset were simulated in our proposed model by shifting relevant PCA components and adjusting speaking rate of the F5-TTS as given in Table~\ref{tab:lombard_params}.

\begin{table}[h]
\centering
\caption{Lombard control parameters used for synthesis.}
\label{tab:lombard_params}

\begin{tabular}{lccc}
\toprule
\textbf{Condition} & $\alpha_E$ & $\alpha_A$ & \textbf{Speed} \\
\midrule
Soft       & -0.5 & -0.5 & 1.0 \\
Normal     &  0.0 &  0.0 & 1.0 \\
Loud       &  0.5 &  0.5 & 0.9 \\
Very Loud  &  1.0 &  1.0 & 0.9 \\
\bottomrule
\end{tabular}

\end{table}

To assess intelligibility robustness to noise, we evaluated the model outputs at four signal-to-noise (SNR) levels: no noise, SNR = 10~dB, SNR = 5~dB, and SNR = 1~dB. We use babble noise as the background noise to simulate a highly masking noisy environment. Importantly, evaluating all conditions at fixed SNR prevents improvements attributed solely to overall signal amplification, which could otherwise be achieved by simply increasing the output volume. This allows us to better assess the contribution of fine-grained acoustic modifications induced by increased vocal effort rather than overall energy gain.

We measure WER to quantify how well the synthesized speech maintains intelligibility under different noise levels. However, absolute WER values in noisy environments could be biased by the model performance on the clean speech samples, which might not have perfect WER scores. Thus, we introduce a noise-robustness metric, relative WER ($\Delta$WER) to achieve a fair evaluation, measuring the WER degradation of noisy audio with respect to the clean audio. The $\Delta$WER defined as normalization of the WER on noisy sample based on the clean sample:

\begin{equation}
\Delta WER = \frac{WER_{\text{noisy}}-WER_{\text{clean}}}{WER_{\text{clean}}}
\end{equation}

Furthermore, to ensure that speaker identity is preserved after simulated Lombard effect, we measured SSIM between synthesized and corresponding ground-truth clean samples at each Lombard level. We also computed the relative SSIM by comparing the  manipulated speech embeddings (soft, loud, very loud) with respect to normal speech, quantifying how consistently speaker identity is maintained across Lombard effect variations.

\begin{table}[t]
\centering
\caption{Comparison of baseline F5-TTS and F5-TTS Style.}

\label{tab:wer_ssim_utmos}
\begin{tabular}{llccc}
\toprule
\textbf{Prompt} & \textbf{Model} & \textbf{WER} & \textbf{SSIM} & \textbf{UTMOS} \\
\midrule
\multirow{2}{*}{English} 
& F5TTS-Base & 2.11 & 95.90 & 3.84 \\
& F5TTS-Style    & 2.08 & 89.10 & 3.53 \\
\midrule
\multirow{2}{*}{German} 
& F5TTS-Base & 7.04 & 96.30 & 3.18 \\
& F5TTS-Style    & 2.48 & 81.50 & 3.49 \\
\bottomrule
\end{tabular}

\end{table}

\subsection{Comparison with Baseline}

Table~\ref{tab:wer_ssim_utmos} compares our proposed F5TTS-Style with the F5TTS-Baseline. On English prompts, F5TTS-Style achieves comparable WER to the baseline, with a slight decrease in UTMOS and speaker similarity. This degradation may be related to the absence of in-context adaptation, as the fixed-size style embedding provides a more compact conditioning representation than the original reference-based pathway.

In the more challenging cross-lingual setting with German prompts, F5TTS-Style substantially outperforms the baseline in both intelligibility (lower WER) and naturalness (higher UTMOS). This improvement suggests that removing in-context reliance and instead using fixed-size style embeddings reduces sensitivity to pronunciation mismatches between languages that can arise from reference-based conditioning. Overall, these results indicate that style embeddings provide more robust performance under evaluated cross-lingual condition.

\begin{figure}[t!]
    \centering
    \includegraphics[width=1.075\linewidth]{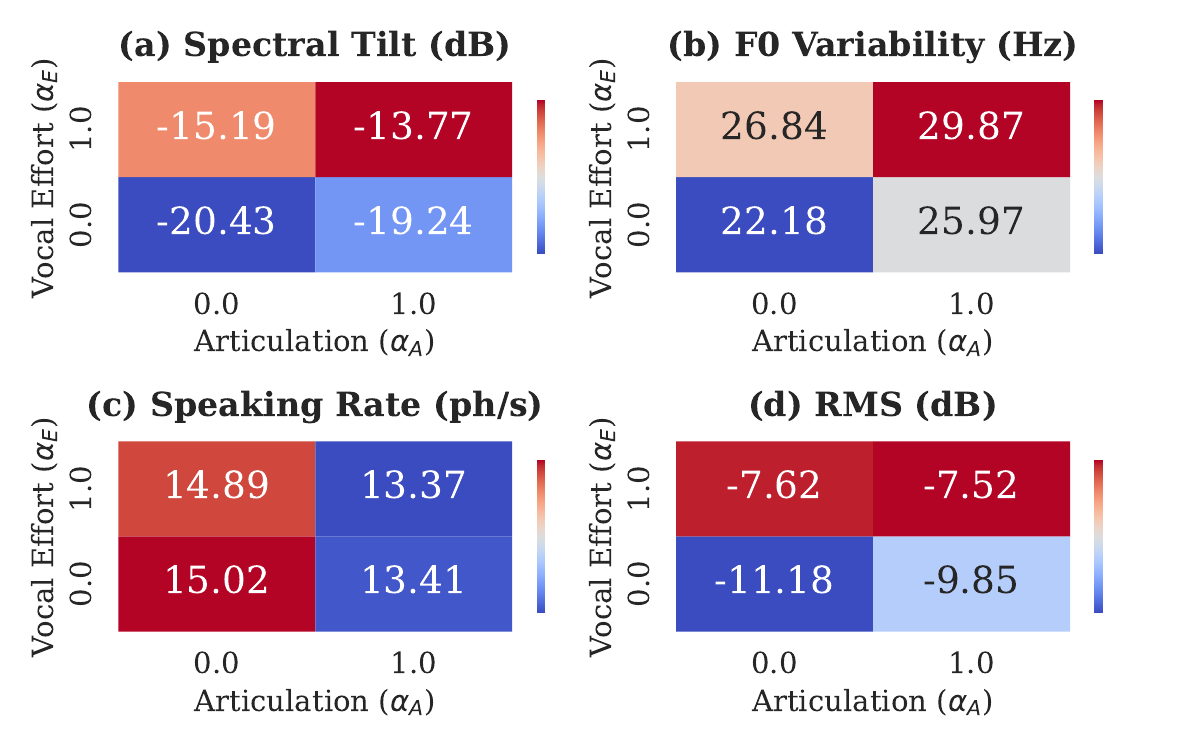}
    \caption{Acoustic changes resulting from shifts along the PCA directions.}
    \label{fig:acoustic_interaction}

\end{figure}

\subsection{Acoustic Effects of PCA-Controlled Lombard Features}
\label{sec:acoustic_effects}
To demonstrate that shifted style embeddings in the identified principal component directions correspond to meaningful signal level changes in the output speech, we report several acoustic metrics:

\begin{itemize}
    \item \textbf{Spectral Tilt:} Spectral tilt is calculated as the ratio of the total energy in higher frequencies (above 5kHz) to the energy in the lower frequencies (1-5 kHz). As vocal effort increases, spectral tilt is expected to get higher (less negative) due to the vocal energy gets shifted to the higher frequencies. As shown in Figure~\ref{fig:acoustic_interaction}(a), spectral tilt primarily increases along the vocal effort axis, while changes along the articulation axis are minimal, indicating largely independent control of vocal effort. 
    \item \textbf{F0 Variability:} F0 variability is measured as the standard deviation of the F0 contour over voiced regions. Lombard speech is generally characterized by an increase in mean F0, together with changes in intensity, duration, and spectral characteristics. Figure~\ref{fig:acoustic_interaction}(b) shows that F0 variability increases along both axes, reflecting contributions from vocal effort and articulation.
    \item  \textbf{Speaking Rate:} Speaking rate is measured as the number of phonemes produced per second and primarily reflects temporal changes associated with articulation. In contrast, vocal effort is expected to have a comparatively smaller effect on speaking rate, with its primary effects expressed through acoustic properties such as intensity and spectral characteristics. This is observed in Figure~\ref{fig:acoustic_interaction}(c), where speaking rate decreases with increasing articulation coefficient, while remaining relatively stable across the vocal effort axis.
    \item \textbf{Root Mean Square Energy (RMS):} RMS reflects overall signal energy, which mainly increases with respect to vocal effort. Figure~\ref{fig:acoustic_interaction}(d) shows that RMS is strongly correlated with the vocal effort axis. There is a minor increase along the articulation axis at moderate vocal effort, indicating that hyper-articulation can slightly boost overall energy due to additional stress on the pronunciation, but at higher vocal effort levels, articulation changes have negligible effect on RMS.
\end{itemize}

\begin{figure}[t]
    \centering

    \begin{subfigure}[t]{0.5\textwidth}
        \centering
        \includegraphics[width=\linewidth]{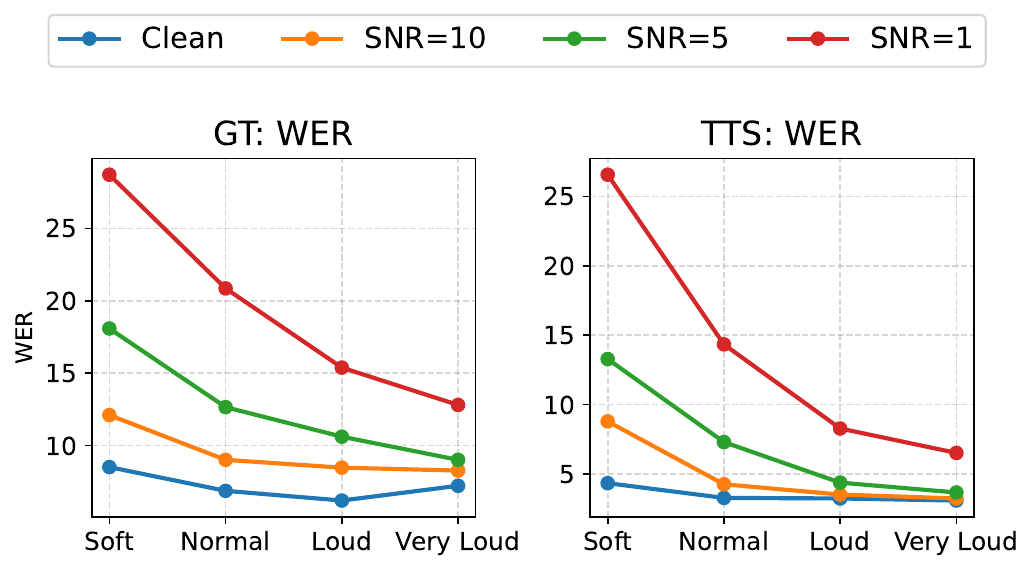}
        \caption{Word Error Rate (WER) under different Lombard levels and noise conditions.}
        \label{fig:wer}
    \end{subfigure}
    \vfill
    \begin{subfigure}[t]{0.5\textwidth}
        \centering
        \includegraphics[width=\linewidth]{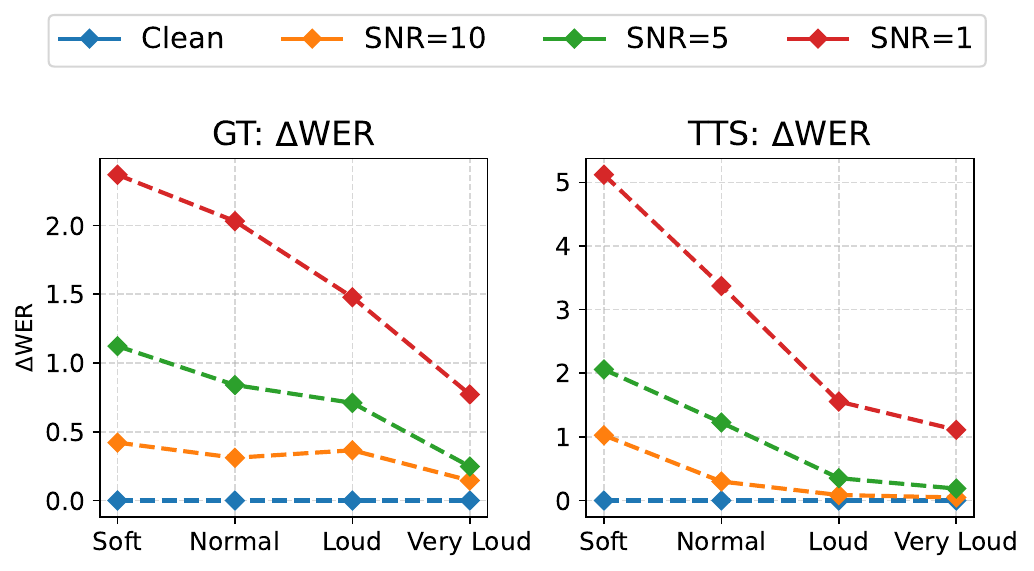}
        \caption{Relative WER degradation ($\Delta$WER) normalized to clean conditions.}
        \label{fig:delta_wer}
    \end{subfigure}

    \caption{Evaluation of intelligibility robustness under Lombard-style speech generation and noise conditions.}
    \label{fig:wer_analysis}

\end{figure}

\subsection{Intelligibility Robustness of Lombard Speech}

Figure~\ref{fig:wer_analysis} illustrates the effect of increasing Lombard level on ASR word error rate (WER) under different noise conditions. As Lombard level increases from soft to very loud, both GT and TTS samples show reduced sensitivity to noise level, with WER curves becoming more compact across SNRs. In particular, $\Delta$WER drops sharply at higher Lombard levels, indicating that intelligibility is preserved even under severe noise (e.g. SNR = 1~dB). These results confirm that manipulating style embeddings to control Lombard effect enhances noise robustness, consistent with the role of Lombard speech in natural communication.

The ablation study results at Table~\ref{tab:wer_ablation} highlights the relative contributions of articulation and vocal effort changes, evaluated in the very loud scenario. When articulation is held constant and only vocal effort is increased (w/o Articulation), WER rises across all noise levels, indicating that articulation is essential for improving intelligibility especially under no noise to moderate noise. However, for the higher noise levels (SNR = 1~dB) dropping articulation has minimal effect. Conversely, when only articulation is increased and vocal effort is unchanged (w/o Vocal Effort), WER remains similar in clean condition but diverges sharply in noisy environments, demonstrating that vocal effort is highly effective under noisy conditions.

\begin{table}[t]
\centering
\caption{Ablation results measured by WER (\%) across different SNR levels in the Very Loud condition.}
\label{tab:wer_ablation}
\resizebox{\linewidth}{!}{
\begin{tabular}{lcccc}
\toprule
\textbf{Condition} & \textbf{No Noise} & \textbf{SNR=10} & \textbf{SNR=5} & \textbf{SNR=1} \\
\midrule
Proposed  & 3.09  & 3.24  & 3.67  & 6.52 \\
\midrule
w/o Articulation & 3.98 & 4.06 & 4.14 & 6.76 \\
w/o Vocal Effort     & 3.01 & 4.02 & 6.29 & 11.99 \\
\bottomrule
\end{tabular}}
\end{table}



\begin{table}[h!]

\centering
\caption{CMOS scores with 95\% confidence interval.}

\label{tab:human_eval}
\resizebox{\linewidth}{!}{

\begin{tabular}{lcc}
\toprule
\textbf{Condition} & \textbf{Naturalness} & \textbf{Intelligibility} \\
\midrule
Clean         & $-0.75 \,\pm\, 0.55$ & $1.48 \,\pm\, 0.38$ \\
Loud (SNR=10) & $-1.29 \,\pm\, 0.46$ & $0.92 \,\pm\, 0.34$ \\
Very Loud (SNR=5) & $-1.60 \,\pm\, 0.33$ & $0.21 \,\pm\, 0.40$ \\
\midrule
Overall       & $-1.22 \,\pm\, 0.26$ & $0.87 \,\pm\, 0.23$ \\
\bottomrule
\end{tabular}}

\end{table}

\textbf{Subjective Evaluation:} We conducted a small-scale user study with 9 participants, assessing three conditions: normal-clean, loud-SNR=10, and very loud-SNR=5, using six samples per condition. Participants listened to pairs of GT and synthesized audio without being informed which sample was GT and rated naturalness and intelligibility. Ratings were given on a [-3, +3] comparative scale, where negative values indicate a preference for GT and positive values indicate a preference for the TTS output. As shown in Table~\ref{tab:human_eval}, GT speech receives higher naturalness ratings, whereas synthesized speech receives higher intelligibility ratings across the evaluated conditions. These results provide preliminary perceptual evidence that the proposed Lombard control can improve perceived intelligibility under selected noisy conditions, while naturalness remains a limitation relative to GT speech.

\subsection{Speaker Identity Preservation in Lombard Speech}

Table~\ref{tab:ssim_results} reports both absolute and relative SSIM scores. Absolute SSIM, measured between ground-truth and synthesized speech, remains consistent across all Lombard levels, indicating that speaker identity is preserved even when the Lombard effect is introduced.

Relative SSIM (R-SSIM), computed by comparing manipulated speech to the corresponding normal synthetic condition, closely follows the trends observed in the ground-truth data. This suggests that the proposed approach enables controllable Lombard synthesis while maintaining stable speaker identity representations.


\begin{table}[h]
\centering
\caption{SSIM results across Lombard levels.}

\label{tab:ssim_results}
\resizebox{\linewidth}{!}{
\begin{tabular}{lcccc}
\toprule
 & \textbf{Soft} & \textbf{Normal} & \textbf{Loud} & \textbf{Very Loud} \\
\midrule
GT vs. TTS  & 80.7\% & 81.2\% & 81.8\% & 81.8\% \\
\midrule
GT R-SSIM       & 93.8\% & --- & \textbf{94.6\%} & \textbf{92.4\%} \\
TTS R-SSIM      & \textbf{94.4\%} & --- & 93.7\% & 91.9\% \\
\bottomrule
\end{tabular}}

\end{table}

\section{Conclusion}
\label{sec:conclusion}

We presented a controllable Lombard TTS framework that enables modulation of vocal effort and articulation through manipulation of learned style embeddings, without requiring Lombard-specific training or fine-tuning of the TTS model. By analyzing the embedding space using PCA and shifting attribute-related components, our approach provides interpretable control over Lombard-related characteristics for speakers not used to estimate the control directions. Experiments show improved recognition robustness under noisy conditions and relatively stable speaker similarity across Lombard control levels, while the cross-lingual evaluation demonstrates improved WER compared with the F5-TTS baseline in the evaluated German-prompt setting. The perceptual study further indicates a trade-off between intelligibility and naturalness, with synthesized speech preferred for intelligibility in the evaluated conditions and ground-truth speech preferred for naturalness.

Overall, our results demonstrate that leveraging large-scale prosodic style representations offers a simple yet effective and scalable approach for zero-shot Lombard speech synthesis also generalizable to unseen speakers. Future work may explore more fine-grained control of prosody and articulation, as well as integration into interactive and real-time speech systems.

\section{Acknowledgments}
This research is supported by the European Union’s Horizon Europe programme grant agreement No. 101213369 (DVPS) and KIT Campus Transfer GmbH (KCT) in accordance with the collaboration with Carnegie-AI.

\bibliographystyle{IEEEtran}
\bibliography{references}
\end{document}